\documentclass{article}
\usepackage{arxiv}

\usepackage[utf8]{inputenc}
\usepackage[T1]{fontenc}
\usepackage{url}
\usepackage{booktabs}
\usepackage{amsfonts}
\usepackage{amsmath}
\usepackage{nicefrac}
\usepackage{microtype}
\usepackage{graphicx}
\usepackage{doi}
\usepackage{verbatim}

\usepackage{placeins}

\usepackage{hyperref}

\title{A Hybrid CAPTCHA Combining Generative AI with Keystroke Dynamics for Enhanced Bot Detection}

\date{}

\author{
 Ayda Aghaei Nia \\
 Institute for Artificial Intelligence\\
 University of Technology\\
 \texttt{aydaaghaeinia@gmail.com} \\
}

\renewcommand{\headeright}{}
\renewcommand{\undertitle}{}
\renewcommand{\shorttitle}{A Hybrid CAPTCHA with Keystroke Dynamics}

\hypersetup{
pdftitle={A Hybrid CAPTCHA Combining Generative AI with Keystroke Dynamics for Enhanced Bot Detection},
pdfsubject={cs.CR, cs.AI},
pdfauthor={Ayda Aghaei Nia}, 
pdfkeywords={CAPTCHA, Bot Detection, Keystroke Dynamics, Generative AI, LLM, Web Security, Behavioral Biometrics},
}

\begin{document}
\maketitle

\begin{abstract}
	Completely Automated Public Turing tests to tell Computers and Humans Apart (CAPTCHAs) are a foundational component of web security, yet traditional implementations suffer from a trade-off between usability and resilience against AI-powered bots. This paper introduces a novel hybrid CAPTCHA system that synergizes the cognitive challenges posed by Large Language Models (LLMs) with the behavioral biometric analysis of keystroke dynamics. Our approach generates dynamic, unpredictable questions that are trivial for humans but non-trivial for automated agents, while simultaneously analyzing the user's typing rhythm to distinguish human patterns from robotic input. We present the system's architecture, formalize the feature extraction methodology for keystroke analysis, and report on an experimental evaluation. The results indicate that our dual-layered approach achieves a high degree of accuracy in bot detection, successfully thwarting both paste-based and script-based simulation attacks, while maintaining a high usability score among human participants. This work demonstrates the potential of combining cognitive and behavioral tests to create a new generation of more secure and user-friendly CAPTCHAs.
\end{abstract}


\section{Introduction}
The proliferation of malicious bots on the internet poses a significant threat to the integrity of online services, ranging from spam and data scraping to denial-of-service attacks. For years, CAPTCHAs have served as the primary defense mechanism \cite{von2003captcha}. However, the rapid advancement of computer vision and machine learning has rendered many traditional CAPTCHAs ineffective. Recent studies have demonstrated that even complex, modern CAPTCHAs can be broken with high accuracy using deep learning models, highlighting the urgent need for new defense paradigms \cite{sivakorn2016im}. This technological arms race often results in more convoluted challenges and a degraded user experience.

In response, systems like Google's reCAPTCHA have shifted towards behavioral analysis, monitoring user interactions to generate a risk score \cite{bursztein2014end}. While effective, these systems often operate as "black boxes" and raise privacy concerns due to extensive user data tracking. This paper explores an alternative paradigm: a transparent, dual-layered CAPTCHA that combines a dynamic cognitive challenge with a localized behavioral biometric analysis.

Our primary contribution is a novel CAPTCHA system that leverages a Generative AI (Large Language Model) to create simple, common-sense questions. This ensures that challenges are unique and not susceptible to replay attacks. Crucially, we augment this cognitive test by analyzing the user's keystroke dynamics—the inherent rhythm of their typing—to provide a powerful second layer of defense. We hypothesize that this hybrid approach can effectively discriminate between humans and bots while offering a more intuitive and less intrusive user experience.

\section{Related Work}
The concept of using automated tests to distinguish humans from computers was formalized by von Ahn et al. \cite{von2003captcha}, leading to the first generation of text-based CAPTCHAs. As these became vulnerable to optical character recognition (OCR) algorithms, subsequent iterations introduced image-based challenges. However, the efficacy of these visual puzzles has been significantly diminished by modern convolutional neural networks \cite{sivakorn2016im}.

More recently, the focus has shifted towards behavioral biometrics. Google's reCAPTCHA v3 is a prominent example, which moves away from explicit challenges entirely, instead relying on a risk analysis engine that processes a wide array of signals captured during a user's session \cite{bursztein2014end}. While powerful, this approach lacks transparency and requires monitoring user behavior across websites, which has significant privacy implications. hCaptcha offers an alternative that focuses on data labeling tasks, but it still relies on manual, puzzle-based interactions.

Keystroke dynamics is a well-established field within behavioral biometrics, primarily used for continuous authentication and user re-verification \cite{monaco2017keystroke}. The rhythm and cadence of typing produce a unique signature that can be used to verify a user's identity. Research in this area has identified numerous features, such as dwell time (duration of a key press) and flight time (time between key presses), that are effective for modeling user-specific patterns \cite{bergadano2002user}.

However, the application of keystroke dynamics as a primary discriminator in a one-shot, public-facing challenge like a CAPTCHA remains under-explored. Most existing work focuses on verifying a known user's identity rather than making a universal human-vs-bot classification. Our work bridges this gap by applying principles of keystroke dynamics analysis in the context of a CAPTCHA and combining it with dynamic, LLM-generated content to create a robust and novel security mechanism.

\section{Methodology}
\subsection{System Design}
The proposed system is designed based on a secure client-server architecture. In this model, the client-side interface is responsible for rendering challenges and capturing user input, while a trusted backend server manages all sensitive operations. This includes making secure API calls to the LLM to generate challenges and performing the final validation of both the user's answer and their behavioral data. This separation of concerns prevents exposure of sensitive information (like API keys or correct answers) on the client side.

\subsection{Dynamic Challenge Generation}
To generate unpredictable challenges, the backend communicates with a Large Language Model (in our implementation, Google's Gemini). The process is guided by structured prompt engineering. A \textbf{system prompt} instructs the LLM to act as a CAPTCHA generator and strictly adhere to a JSON output format. A randomized \textbf{user prompt} (e.g., "Ask a simple question about colors") ensures variety.
\begin{verbatim}
// Example LLM JSON Response
{
  "question": "What color is the sky on a clear day?",
  "answer": "blue"
}
\end{verbatim}
Upon receiving the response, the backend computes the SHA-256 hash of the answer and transmits only the question and the hash to the client, ensuring the plaintext answer is never exposed.

\begin{figure}[htbp]
	\centering
	\includegraphics[width=0.6\textwidth]{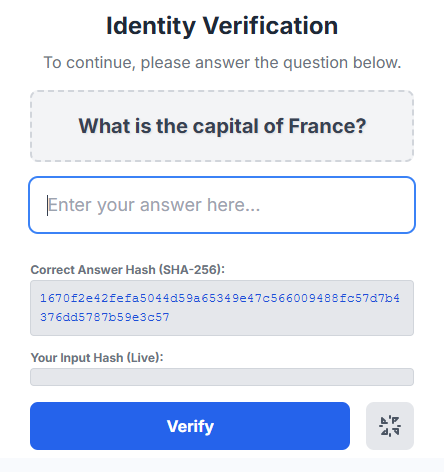}
	\caption{A general knowledge question generated by the LLM.}
	\label{fig:text_challenge}
\end{figure}

\begin{figure}[htbp]
	\centering
	\includegraphics[width=0.6\textwidth]{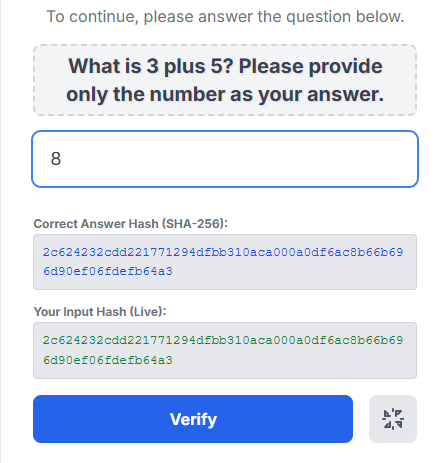}
	\caption{A mathematical question to demonstrate challenge variety.}
	\label{fig:math_challenge}
\end{figure}

\subsection{Behavioral Feature Extraction}
When the user types their answer, the client-side script captures high-precision timestamps for each `keydown` event using `performance.now()`. From this raw data, we extract a set of behavioral features.

Let the sequence of key press timestamps be $T = \{t_1, t_2, ..., t_n\}$ for an $n$-character input.
We define the inter-key latency, or \textbf{flight time}, $F_i$, between consecutive key presses as:
\begin{equation}
	F_i = t_{i+1} - t_i, \quad \text{for } i = 1, ..., n-1
\end{equation}
From the sequence of flight times $F = \{F_1, F_2, ..., F_{n-1}\}$, we compute three statistical metrics:
\begin{enumerate}
	\item \textbf{Total Typing Duration ($T_{total}$):} The total time elapsed from the first to the last keystroke. $T_{total} = t_n - t_1$.
	\item \textbf{Mean Latency ($\mu_F$):} The average time between key presses.
	\item \textbf{Standard Deviation of Latency ($\sigma_F$):} The variability in the typing rhythm. This is a key feature, as human typing is inherently less uniform than that of simple scripts.
\end{enumerate}

\subsection{Classification Heuristic}
A user's input is classified as \textbf{human} if and only if all of the following conditions are met:
\begin{enumerate}
    \item The SHA-256 hash of the typed answer matches the target hash.
    \item A browser `paste` event was not detected.
    \item The extracted behavioral metrics fall within empirically determined thresholds. Specifically, the standard deviation of latency must be above a minimum threshold ($\sigma_F > \theta_{\sigma}$) to rule out robotic uniformity, and the total time must be above a threshold ($T_{total} > \theta_{t}$) to prevent high-speed inputs.
\end{enumerate}
If any of these conditions are not met, the input is classified as \textbf{bot}.

\section{Experimental Evaluation}
\subsection{Experimental Setup}
The system was evaluated in a modern web browser environment (Google Chrome). Timestamps were captured using the JavaScript `performance.now()` API for sub-millisecond precision. Based on preliminary empirical testing, the classification thresholds were set to $\theta_{\sigma} = 20ms$ for the standard deviation of latency and $\theta_{t} = 150ms$ for the total duration of answers with more than three characters.

Two participant groups were used:
\begin{itemize}
    \item \textbf{Human Group:} 15 volunteers were asked to solve 3 CAPTCHA challenges each.
    \item \textbf{Bot Group:} Two automated scripts were developed using a Python library. The first was a \textbf{paste-based bot} that directly inserted the correct answer. The second was a \textbf{typing-simulation bot} that entered the answer character-by-character with a fixed 50ms delay. Each bot was run in 50 trials.
\end{itemize}

\subsection{Results}
The results of the evaluation are summarized in Table \ref{tab:results}. The system demonstrated high usability for human users, with a first-attempt success rate of 87\% (total success rate of 100\% within two attempts). The primary reason for initial failure was typographical errors, which is expected.

Crucially, the system achieved a 100\% detection rate against both bot types. The paste-based bot was consistently blocked by the event listener, while the typing-simulation bot was consistently flagged by the behavioral analysis due to its unnaturally low latency variation ($\sigma_F \approx 0$), failing the $\sigma_F > \theta_{\sigma}$ check.

\begin{table}[htbp]
	\caption{Experimental Evaluation Results}
	\centering
	\begin{tabular}{l c c l}
		\toprule
		\textbf{Participant Group} & \textbf{Trials} & \textbf{Success Rate (\%)} & \textbf{Primary Reason for Failure} \\
		\midrule
		Human Users & 45 & 100 & Typographical Errors (on 1st attempt) \\
		Bot (Paste-based) & 50 & 0 & Paste event detected \\
		Bot (Typing Simulation) & 50 & 0 & Low latency std. deviation ($\sigma_F < \theta_{\sigma}$) \\
		\bottomrule
	\end{tabular}
	\label{tab:results}
\end{table}

\begin{figure}[htbp]
	\centering
	\includegraphics[width=0.6\textwidth]{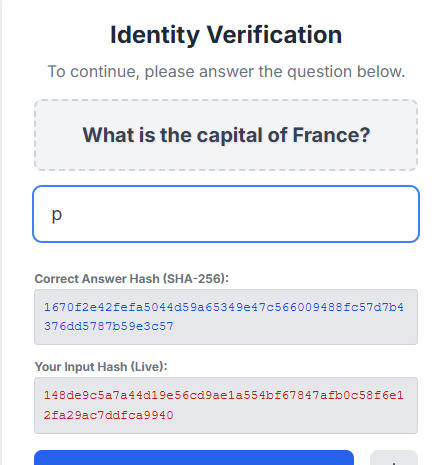}
	\caption{Real-time hashing shows the user's input hash (bottom) does not yet match the correct answer hash (top).}
	\label{fig:mismatched_hash}
\end{figure}

\begin{figure}[htbp]
	\centering
	\includegraphics[width=0.6\textwidth]{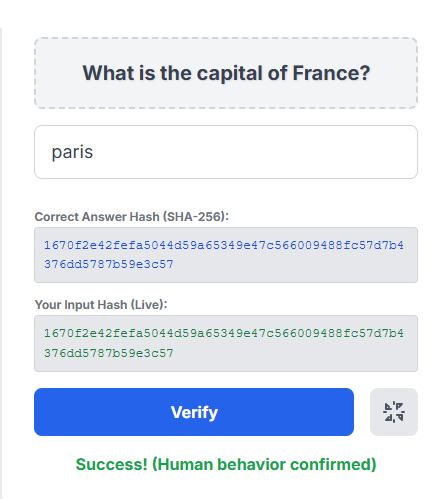}
	\caption{A successful verification message is displayed after the correct answer is typed with human-like behavior.}
	\label{fig:success}
\end{figure}

\section{Discussion}
The experimental results strongly support our initial hypothesis. The dual-layered approach of combining a cognitive challenge with behavioral biometric analysis proved highly effective. The LLM-based questions ensure semantic novelty, preventing replay attacks, while the keystroke dynamics analysis successfully filtered out non-human interaction patterns. The high usability score further suggests that this model is a viable alternative to more intrusive or cognitively demanding CAPTCHAs.

However, we acknowledge several limitations. The primary limitation is the simplistic nature of our bot adversary. A more sophisticated bot could be programmed to inject randomized delays between keystrokes to mimic human variability. Defending against such an adversary would require a more advanced classification model than our current heuristic-based approach. Secondly, our evaluation was conducted with a small user population; a larger, more diverse dataset is needed to validate the generalizability of the chosen thresholds. Finally, the system relies on an external LLM API, which introduces dependencies related to cost, latency, and service availability.

\section{Conclusion and Future Work}
In this paper, we presented a novel hybrid CAPTCHA system that integrates generative AI with keystroke dynamics. Our proof-of-concept demonstrates that this approach is highly effective at detecting simple bots while providing a seamless and positive user experience. By shifting the security focus from "what you know" to "what you know and how you act," we can build more robust and humane verification systems.

Future work will proceed along three main avenues:
\begin{enumerate}
    \item \textbf{Advanced Behavioral Modeling:} We plan to replace the current heuristic classifier with a machine learning model, such as a one-class Support Vector Machine (SVM) or an autoencoder. By training on a large dataset of human typing patterns, these models can create a more nuanced and robust classification boundary, making it significantly harder for sophisticated bots to evade detection.
    \item \textbf{Multi-modal Biometrics:} Keystroke dynamics can be augmented with other behavioral signals. We will investigate incorporating mouse movement analysis, tracking features like cursor velocity, acceleration, and path curvature to build a richer, multi-modal behavioral profile.
    \item \textbf{Performance and Scalability Analysis:} A comprehensive analysis of the system's performance under load is required. This includes studying the latency overhead of the LLM API calls and optimizing the client-server communication protocol for a production environment.
\end{enumerate}


\begin{thebibliography}{9}

\bibitem{von2003captcha}
Luis Von Ahn, Manuel Blum, Nicholas J. Hopper, and John Langford.
\newblock CAPTCHA: Using hard AI problems for security.
\newblock In {\em International Conference on the Theory and Applications of Cryptographic Techniques (EUROCRYPT)}, pages 294--311. Springer, 2003.

\bibitem{bursztein2014end}
Elie Bursztein, Adam Pritchard, Arel Cordero, and Matthieu Martin.
\newblock The end is nigh: generic solving of text-based CAPTCHAs.
\newblock In {\em 8th USENIX Workshop on Offensive Technologies (WOOT)}, 2014.

\bibitem{sivakorn2016im}
Suphannee Sivakorn, Jason Polakis, and Angelos D. Keromytis.
\newblock I'm not a human: Breaking the Google reCAPTCHA.
\newblock In {\em 2016 IEEE European Symposium on Security and Privacy (EuroS\&P)}, pages 191--205. IEEE, 2016.

\bibitem{monaco2017keystroke}
J. V. Monaco, J. T. Monaco, C. C. Tappert, et al.
\newblock Keystroke biometric systems for user authentication.
\newblock In {\em 2017 IEEE International Carnahan Conference on Security Technology (ICCST)}, pages 1--8. IEEE, 2017.

\bibitem{bergadano2002user}
Francesco Bergadano, Daniele Gunetti, and Claudia Picardi.
\newblock User authentication through keystroke dynamics.
\newblock {\em ACM Transactions on Information and System Security (TISSEC)}, 5(4):367--397, 2002.

\end{thebibliography}
\end{document}